\documentclass[prl,aps,preprint]{revtex4}
\usepackage[]{times,amsmath,epsfig,graphicx}
\DeclareRobustCommand{\baselinestretch{2}}

\begin{document}

\title{Whispering gallery mode resonator based ultra-narrow linewidth external cavity semiconductor laser}

\author{W. Liang, V. S. Ilchenko, A. A. Savchenkov, A. B. Matsko, D. Seidel, and L. Maleki}

\affiliation{OEwaves Inc., 2555 E. Colorado Blvd., Ste. 400, Pasadena, California 91107}

\begin{abstract}
We demonstrate a miniature self-injection locked DFB laser using resonant optical feedback from a high-Q crystalline whispering gallery mode resonator. The linewidth reduction factor is greater than 10,000, with resultant instantaneous linewidth less than 200~Hz. The minimal value of the Allan deviation for the laser frequency stability is $3\times 10^{-12}$ at the integration time of 20~$\mu$s. The laser possesses excellent spectral purity and good long term stability.
\\ OCIS: 140.3425, 140.3520, 140.2020, 140.4780, 290.5870
\end{abstract}

\maketitle

Highly coherent, narrow linewidth lasers with long term frequency stability have applications in many areas of optics including spectroscopy, metrology, remote sensing, communications, and bio-chemical sensing. Semiconductor lasers stabilized with external cavities are readily available, but only in a few spectral intervals corresponding to optical communications bands. Modern semiconductor lasers allow coverage of nearly the entire optical spectrum, though their coherence properties are frequently inadequate for many applications. To improve the spectral properties and generate narrow linewidth light over a broad range of wavelengths, we use self-injection locking \cite{dahmani87ol,hollberg88apl,himmerich94ao} with high quality factor (Q) whispering gallery mode (WGM) resonators.

WGM resonators are useful for laser stabilization since they provide  very high-Q in a broad wavelength range. Self-injection locking \cite{dahmani87ol,hollberg88apl,himmerich94ao} is one of the most efficient ways to lock a laser to a WGM. The method is based on  resonant Rayleigh scattering in the resonator \cite{gorodetsky00josab}: as a result of scattering some amount of light reflects back into the laser when the frequency of the emitted light coincides with the frequency of the selected WGM. This provides a very fast optical feedback enabling a significant reduction of  laser linewidth. Several experiments on self-injection locking of various lasers to a WGM resonator have been previously reported \cite{vassiliev98oc,vassiliev03apb,kieu07ol,spengler09ol}.

Self-injection locking to a WGM is applicable to any laser emitting at the wavelength within the transparency window of the resonator host material. For instance, lasers emitting in the 150~nm-10~$\mu$m range can be stabilized using CaF$_2$ WGM resonators. This leads to a broad range of opportunities for realizing miniature narrow-line lasers suitable for any application where low optical phase and frequency noise  is important. As the self-injection locking does not require any electronics, the laser can be very tightly packaged, which simplifies its thermal stabilization, and also reduces the influence of the acoustic noise on the laser frequency. The laser can be used as a master laser for pumping high power lasers used for metrology and remote sensing.

This approach for self injection locking also produces a tunable semiconductor laser. A widely tunable laser locked to a WGM resonator can be obtained with temperature tuning. The tuning speed, however, is comparably slow in this case and is determined by the thermal response of the resonator fixture, and the tuning rate is usually in the range of several gigahertz per degree. An agile frequency tuning of  the laser can be achieved by other means. For instance, electro-optic resonators with voltage controlled spectra can be used to injection lock the laser. The tuning agility in such a system is determined by the ring down time of the WGM mode, which can be shorter than a microsecond.

Here we describe results of our experiments on self injection locking of DFB diode lasers using crystalline WGM resonators. We have achieved instantaneous linewidth of less than 200~Hz in such lasers, with long frequency stability limited only by the thermal drift of the WGM frequency. While our investigations were performed with lasers at telecommunication wavelength, they can be considered as a proof of principle for locking lasers at any wavelength where the WGM resonator has high enough Q-factor.

\begin{figure}[ht]
\centerline{\epsfig{file=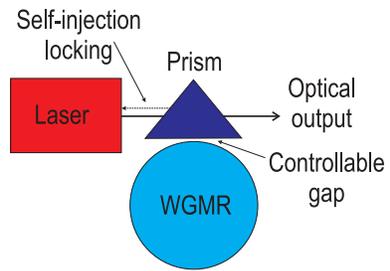,width=5.0cm,angle=0}}
\caption{\label{figure1}  W. Liang et al., "Whispering gallery mode resonator based ultra-narrow linewidth external cavity semiconductor laser"}
\end{figure}

The laser setup is schematically shown in Fig.~(\ref{figure1}). Light emitted by a 1550~nm DFB laser mounted on a ceramic submount is collimated and sent into a CaF$_2$ WGM resonator using a coupling prism. The power at the output of the laser chip is 5~mW. The power at the exit of the prism is approximately 3~mW (the reason the power exiting the prism is less than that of  the free running laser is because of the absorption of the resonator when it is nearly critically coupled). The free running laser has 2~MHz instantaneous linewidth.

The resonator is 2~mm in diameter with an unloaded Q-factor approaching $2\times 10^9$. The prism coupler loads the mode until the Q approaches $1\times10^9$. Strong surface Rayleigh scattering \cite{gorodetsky00josab}, responsible for the self-injection locking, results in forming WGM doublets with frequency splitting on the order of 100~kHz.

A significant amount of light reflects back into the laser as a result of the scattering, leading to the lock of the laser frequency to the WGM. Following the WGM frequency, the laser frequency can be pulled in a excess of 4~GHz, leading to several mA of locking range (in the laser current units). The laser and the resonator are mounted on a common thermal control element, and the thermal drift of the WGM (being on the order of a couple of MHz) determines the long term stability of the laser. Another source of drift is related to the heating of the resonator due to absorption of light. A drift of the frequency of free running laser moves the locking point, which changes the circulation power and hence the temperature of the resonator.

To characterize the phase and frequency noise of a laser operating far from threshold we recall that the quantum noise limited linewidth is given by $\Delta \nu = 2 \pi^2 \hbar \nu_0 (\Delta \nu_c) ^2/P$, where $\Delta \nu_c$, an equivalent to FWHM of the laser cavity, varies depending on the laser type and linewidth broadening factor $\alpha$, $\nu_0$ is the carrier frequency, $P$ is the output power. It is worth noting that the linewidth is the FWHM of the Lorentzian spectrum of the laser, and that $\Delta \nu$ is two times smaller as compared with the value of Schawlow-Townes linewidth. The phase diffusion associated with the linewidth is given by $\langle (\Delta \phi (\tau))^2 \rangle = 2 \pi \Delta \nu \tau$.

The single side power density of the phase noise is $ {\cal L}(f)\left [ Hz^{-1} \right ]= \Delta \nu/(2 \pi f^2)$, and the corresponding frequency noise is $S_\nu(f)\left [ Hz^{2}/Hz \right ]=2 f^2 {\cal L}(f)$ ($\Delta \nu = \pi S_\nu$). The Allan variance of the frequency of an ideal laser is given by
\begin{equation} \label{sigm}
\sigma^2(\tau)=2 \int \limits_{0}^{\infty} \frac{S_\nu}{\nu_0^2} \frac{\sin^4(\pi f \tau)}{(\pi f \tau)^2} df= \frac{\Delta \nu}{\nu_0} \frac{1}{\nu_0 \tau}.
\end{equation}

The model is not applicable to a generic laser suffering from $1/f$-noise, so the following expression is used to determine an effective linewidth $\Delta \nu_{eff}$ of the laser  \cite{hjelme91jqe}
\begin{equation} \label{linewidth}
\int \limits_{\Delta \nu_{eff}}^{\infty} {\cal L}(f) df=\frac{1}{2 \pi} \;\;{\rm [rad^2]},
\end{equation}
which gives $\Delta \nu_{eff}=\Delta \nu$ in the case of an ideal laser.

We built two identical lasers and beat their emission on a fast photodiode to study the noise properties of the self injection locked laser. By changing the temperature of one of the lasers we shifted its frequency to be approximately 10~GHz away from the frequency of the second laser. The RF signal generated on the photodiode was used to analyze the spectral properties of the lasers.

Light from two lasers each with power $P/2$  and carrier frequencies $\nu_1$ and $\nu_2$ ($\nu_{1,2} \approx \nu_0$) produces an RF signal with carrier frequency $|\nu_1-\nu_2|$ on a fast photodiode. The signal is characterized with phase noise
\begin{equation} \label{beat1}
{\cal L}_{beat}(f)= \frac{4\pi \hbar \nu_0}{\eta a P} \left ( 1+\eta a \frac{(\Delta \nu_c)^2}{f^2} \right )+ \frac{Fk_BT}{\rho {\cal R}^2 a^2 P^2}+ RIN
\end{equation}
in the case of lasers with quantum limited linewidth, where $F$ is the noise figure of the RF circuit, $k_B$ is the Boltzmann constant, $T$ is the external temperature, $\eta$ is the quantum efficiency of the photodiode, $a$ is the attenuation factor, $\rho$ and ${\cal R}=q \eta/2\pi \hbar \nu_0 $ are the resistance and responsivity of the photodiode, respectively, $q$ is the electron charge, $RIN$ is the relative intensity noise of the lasers. Naturally, the frequency dependent term in Eq.~(\ref{beat1}) describes the laser linewidth, which is not influenced by the signal loss or the efficiency of the photodiode. In the case of beating two typical lasers on a photodiode the phase noise of the RF signal, ${\cal L}_{beat}(f)$, contains $f^{-l}$ terms ($l=1,\dots,4$). Substituting those terms into Eq.~(\ref{linewidth}) we find $\Delta \nu_{eff}$ for the linewidth of the beat note. In some cases flicker noise of the photodiode may need to be taken into account.

\begin{figure}[ht]
\centerline{\epsfig{file=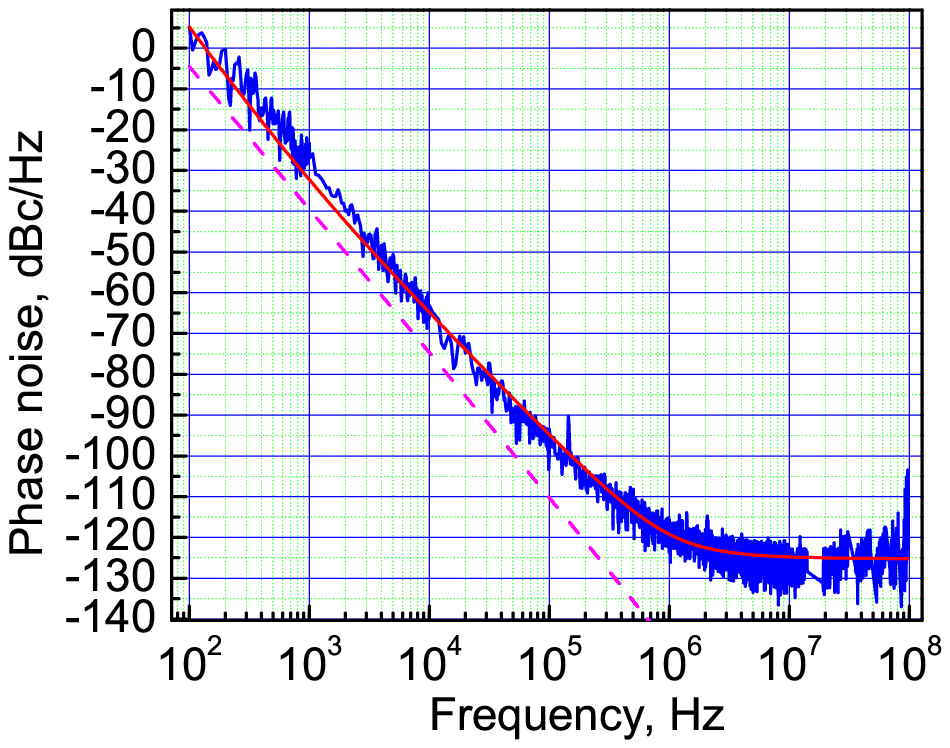,width=6.5cm,angle=0}}
\caption{\label{figure2} W. Liang et al., "Whispering gallery mode resonator based ultra-narrow linewidth external cavity semiconductor laser"}
\end{figure}

An example of the phase noise of the laser beat note is shown in Fig.~(\ref{figure2}). The beat frequency was on the order of $7-10$~GHz, depending on the temperature difference of the two resonators. The power of lasers on the photodiode were about 0~dBm each. A 20~GHz photodiode (Discovery DSC720) was utilized in the measurement. We used a commercial measurement system developed by OEwaves (OE8000) to measure the phase noise. We compared the measured noise with the fundamental thermodynamics-limited phase noise \cite{matsko07josab} and found that the system stability is restricted by the technical noise at a level 10~dB away from the fundamental noise.

It is reasonable to find the linewidth of the beat note in accordance with Eq.~(\ref{linewidth}). To do this we fitted the phase noise dependence using the decomposition over frequency and calculated the integral. The resultant linewidth is 1.2~kHz. Hence, a single laser has sub-kHz linewidth.  The phase noise of the beat note at low frequencies has a signature of thermal drift.

\begin{figure}[ht]
\centerline{\epsfig{file=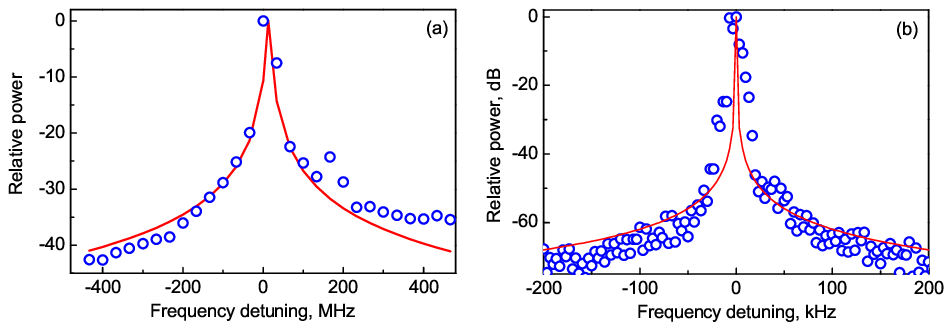,width=8.5cm,angle=0}}
\caption{\label{figure3} W. Liang et al., "Whispering gallery mode resonator based ultra-narrow linewidth external cavity semiconductor laser"}
\end{figure}

We studied the RF beat note using an RF spectrum analyzer to determine the instantaneous linewidth of each laser. Drift of the beat note did not allow us to reduce the resolution bandwidth significantly. We used 18~kHz resolution bandwidth and fitted the skirt of the line with a Lorentzian curve of 160~Hz full width at half maximum (FWHM), shown in Fig.~(\ref{figure3}b). This indicates \cite{vassiliev98oc} that the instantaneous linewidth of the individual laser is less than 160~Hz. To compare the linewidth of the locked and unlocked lasers, we have measured the linewidth of the free running lasers using the same technique and obtained 8~MHz FWHM value for the beat note (Fig.~\ref{figure3}a), which corresponds to the laser specifications supplied by the manufacturer.

\begin{figure}[ht]
\centerline{\epsfig{file=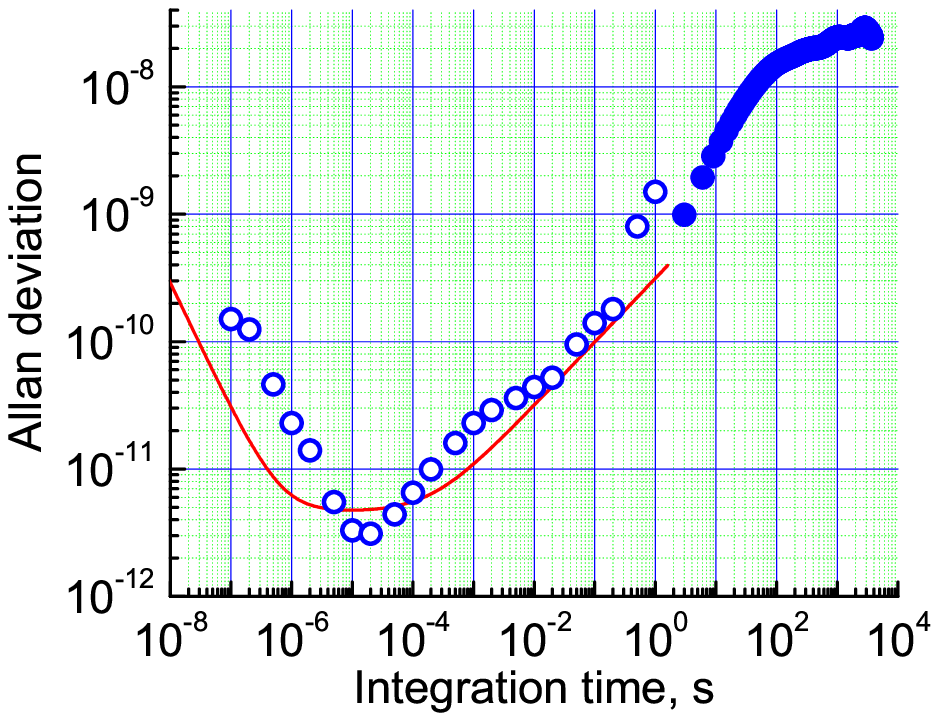,width=6.5cm,angle=0}}
\caption{\label{figure4} W. Liang et al., "Whispering gallery mode resonator based ultra-narrow linewidth external cavity semiconductor laser"}
\end{figure}

Finally, we measured the Allan deviation of the RF beat note directly using the frequency counter technique, and recalculated the data to obtain the Allan deviation of the optical signal. Results are shown in Fig.~(\ref{figure4}). The smallest value of the Allan deviation corresponds to 600~Hz, which has the same order of magnitude as the instantaneous linewidth. To verify the measurement, we took the fit of the phase noise  data (Fig.~\ref{figure2}) and calculated the short term Allan deviation using Eq.~(\ref{sigm}). The resultant curve agrees well with the measurement results.

To conclude, we have demonstrated narrow linewidth DFB lasers self-injection locked with means of WGM resonators.  While these first prototypes had dimensions of about 20 cubic centimeters, the device can be realized in a  package with 15 mm sides with a thickness less than 3 mm. These lasers have good short and long term stability, with instantaneous linewidth smaller than 200~Hz; their long term frequency drift is less than 10~MHz. Narrow linewidth and frequency stable semiconductor lasers of this type are important for various applications in sensing and metrology. Our experiments indirectly show that any DFB laser at any wavelength can be locked to a WGM resonator using the same principle, especially at wavelengths where lasers with narrow linewidth are currently not accessible. Other kinds of semiconductor lasers may also be used instead of the DFB structures.

The authors acknowledge expert help from Danny Eliyahu with the phase noise measurement.



\vspace{24.cm}

{\bf Figure captions:}

\begin{enumerate}

\item[Figure 1]
Schematic of the experimental setup. Light from the pump laser enters the whispering gallery mode resonator (WGMR) through the prism. Part of light is reflected back to the laser due to Rayleigh scattering in the resonator. The light exiting the prism is collimated and sent to an optical spectrum analyzer.

\item[Figure 2]
Single-sided phase noise spectrum of the RF beat signal generated by two self injection locked DFB lasers on a fast photodiode. The solid line  is the fit of the noise using decomposition of terms $f^{-l}$ ($l=1,\dots,4$). The dashed line is the fundamental limit of the phase noise determined by the thermodynamic fluctuation of the WGM frequencies. Interestingly, the low frequency phase noise has $f^{-7/2}$ frequency dependence, similarly with the theoretical limit.

\item[Figure 3]
Frequency spectrum of the RF signal generated by beating two DFB lasers on a fast photodiode. (a) The lasers are free running. The skirts of the line are fitted with 8~MHz Lorentzian envelop. (b) The lasers are self-injection locked. Line shape taken with 18~kHz resolution bandwidth. The line is inhomogeneously broadened due to the frequency drift. The skirts of the line are fitted with 160~Hz Lorentzian envelope.

\item[Figure 4]
Allan deviation of the RF signal produced by beating two self-injection locked DFB lasers on a fast photodiode. We used a signal analyzer to measure the short term stability (empty circles), and a frequency counter to measure the long term stability (solid circles). The solid line represents the curve calculated using the fitting line in Fig.~(\ref{figure2}).

\end{enumerate}

\newpage

\newpage

\newpage

\newpage

\begin{thebibliography}{99}


\bibitem{dahmani87ol} B. Dahmani, L. Hollberg, and R. Drullinger, "Frequency stabilization of semiconductor lasers by resonant optical feedback," Opt. Lett. {\bf 12}, 876-878 (1987).

\bibitem{hollberg88apl} L. Hollberg and M. Ohtsu, "Modulatable narrow-linewidth semiconductor lasers," Appl. Phys. Lett. {\bf 53}, 944-946 (1988).

\bibitem{himmerich94ao} A. Hemmerich, C. Zimmermann, and T. W. Ha\"nsch, "Compact source of coherent blue light," Appl. Opt. {\bf 33}, 988-991 (1994).

\bibitem{gorodetsky00josab} M. L. Gorodetsky, A. D. Pryamikov, and V. S. Ilchenko, "Rayleigh scattering in high-Q microspheres," J. Opt. Soc. Am. B {\bf 17}, 1051-1057 (2000).

\bibitem{vassiliev98oc} V. V. Vassiliev, V. L. Velichansky, V. S. Ilchenko, M. L. Gorodetsky, L. Hollberg, and A. V. Yarovitsky, "Narrow-line-width diode laser with a high-Q microsphere  resonator", Opt. Comm. {\bf 158}, 305-312 (1998).

\bibitem{vassiliev03apb} V. V. Vassiliev, S. M. Ilina, and V. L. Velichansky, "Diode laser coupled to a high-Q microcavity via a GRIN lens", Appl. Phys. B {\bf 76}, 521-523 (2003).

\bibitem{kieu07ol} K. Kieu and M. Mansuripur, "Fiber laser using a microsphere resonator as a feedback element," Opt. Lett. {\bf 32}, 244-246 (2007).

\bibitem{spengler09ol} B. Sprenger, H. G. L. Schwefel, and L. J. Wang, "Whispering-gallery-mode-resonator-stabilized narrow-linewidth fiber loop laser," Opt. Lett. {\bf 34}, 3370-3372 (2009).

\bibitem{hjelme91jqe} D. R. Hjelme, A. R. Mickelson, and R. G. Beausoleil, "Semiconductor laser stabilization by external feedback," IEEE J. Quantum Electron. {\bf 27}, 352-372 (1991).

\bibitem{matsko07josab} A. B. Matsko, A. A. Savchenkov, N. Yu, and L. Maleki, "Whispering-gallery-mode resonators as frequency references. I. Fundamental limitations," J. Opt. Soc. Am. B {\bf 24}, 1324-1335 (2007).


\end{thebibliography}
\end{document}